\documentclass[pre,superscriptaddress,showpacs,eqsecnum,twocolumn]{revtex4}
\usepackage[dvips]{graphicx}
\usepackage{color}
\usepackage{amsmath}
\usepackage{time}
\newcommand{\be}{\begin{equation}}
\newcommand{\ee}{\end{equation}}
\newcommand{\bea}{\begin{eqnarray}}
\newcommand{\eea}{\end{eqnarray}}
\newcommand{\sn}{{\rm sn}}
\newcommand{\ds}{{\rm ds}}
\newcommand{\cs}{{\rm cs}}
\newcommand{\ns}{{\rm ns}}
\newcommand{\dn}{{\rm dn}}
\newcommand{\cn}{{\rm cn}}
\newcommand{\sech}{{\rm sech}}
\preprint{LA-UR-05-1117}
\begin{document}
\title{Exact solitary wave solutions  for a discrete $\lambda \phi^4$
field theory in $1+1$ dimensions}
\author{Fred Cooper} \email{fcooper@nsf.gov}
\affiliation{National Science Foundation,
   Division of Physics,
   Arlington, VA 22230}
\affiliation{Theoretical Division,
   Los Alamos National Laboratory,
   Los Alamos, NM 87545}
\affiliation{Santa Fe Institute,
   Santa Fe, NM 87501}
\author{Avinash Khare} \email{khare@iopb.res.in}
\affiliation{Institute of Physics,
   Bhubaneswar, Orissa 751005, India}
\author{Bogdan Mihaila} \email{bmihaila@lanl.gov}
\affiliation{Theoretical Division,
   Los Alamos National Laboratory,
   Los Alamos, NM 87545}
\author{Avadh Saxena}  \email{avadh@lanl.gov}
\affiliation{Theoretical Division,
   Los Alamos National Laboratory,
   Los Alamos, NM 87545}
\date{\today, \now}
\begin{abstract}
We have found exact, periodic, time-dependent solitary wave
solutions of a discrete $\phi^4$ field theory model. For finite
lattices, depending on whether one is considering a repulsive or
attractive case, the solutions are either Jacobi elliptic functions
$\sn(x,m)$ (which reduce to the kink function $\tanh(x)$ for
$m\rightarrow 1$), or they are $\dn(x,m)$ and $\cn(x,m)$ (which
reduce to the pulse function $\sech(x)$ for $m\rightarrow 1$). We
have studied the stability of these solutions numerically, and we
find that our solutions are linearly stable in most cases. We show
that this model is a Hamiltonian system, and that the effective
Peierls-Nabarro barrier due to discreteness is zero not only for
the two localized modes but even for all three periodic solutions.
We also present results of numerical simulations of scattering of
kink--anti-kink and pulse--anti-pulse solitary wave solutions.
\end{abstract}


\maketitle

\section{Introduction}

Discrete nonlinear equations are ubiquitous and play an important
role in diverse physical contexts~\cite{clark,scott}.  Some examples
of integrable discrete equations include Ablowitz-Ladik (AL)
lattice~\cite{AL} and Toda lattice~\cite{toda}. Certain non-Abelian
discrete integrable models are also known \cite{nonabel}.  There are
many non-integrable discrete nonlinear equations such as discrete
sine-Gordon (DSG)~\cite{DSG1,DSG2}, discrete nonlinear Schr\"odinger
(DNLS) equation~\cite{DNLS} and the Fermi-Pasta-Ulam (FPU)
problem~\cite{ford}. DSG is a physical realization of the dynamics
of dislocations in crystals where it is known as the Frenkel-Kontorova
model~\cite{nabarro}.  It also arises in the context of ferromagnets
with planar anisotropy~\cite{mikeska}, adsorption on a crystal lattice
and pinned charge-density waves~\cite{sacco}.  Similarly, DNLS plays
a role in the propagation of electromagnetic waves in doped glass
fibers~\cite{fibers} and other optical waveguides~\cite{OP}, and it
describes Bose-Einstein condensates in optical lattices~\cite{ST}.
FPU has served as a fertile paradigm for understanding solitons,
discrete breathers, intrinsic localized modes, chaos, anomalous
transport in low-dimensional systems and the fundamentals of
statistical mechanics~\cite{ford}.  A discrete double well or
discrete $\phi^4$ equation is a model for structural phase
transitions~\cite{dphi4}, and may be relevant for a better
understanding of the collisions of relativistic kinks~\cite{luis}.

Obtaining exact (soliton-like) solutions is always desirable,
particularly for discrete systems where the notion of a discreteness
(or Peierls-Nabarro) barrier~\cite{PN,peyrard} is an important one
in that its absence is a likely indication of integrability, e.g. in
the AL case~\cite{scott}.  In addition, exact solutions allow one
to calculate certain important physical quantities analytically as
well as they serve as diagnostics for simulations.  Recently derived
summation identities~\cite{khare} involving Jacobi elliptic
functions~\cite{stegun,byrd} led to exact periodic solutions of a
modified DNLS equation~\cite{dnls}.  In this paper we exploit similar
identities to obtain exact periodic solutions of the discrete
$\phi^4$ model.

The paper is organized as follows. In the next section we summarize
the exact solitary wave solutions of the continuum $\phi^4$ model.
In Sec.~\ref{sec:discrete}, by identifying the relevant elliptic
function identities, we derive a discrete $\phi^4$ model which
allows for exact solutions. We then obtain these solutions and study
their stability. In Sec.~\ref{sec:hdyn} we explicitly write down the
corresponding Hamiltonian dynamics using a modified Poisson bracket
algebra.  Sec.~\ref{sec:scat} contains numerical results for the
scattering of both kink- and pulse-like solitary waves.
Kink--anti-kink collisions appear to create a breather with some
radiation, and pulse--anti-pulse collisions lead to a flip and
little radiation. In Sec.~\ref{sec:PN} we compute the energy of the
solitary waves, and show that the Peierls-Nabarro
barrier~\cite{PN,peyrard} due to discreteness is zero not only for
the two localized solutions but even for the three periodic
solutions. Finally, we summarize our main findings in
Sec.~\ref{sec:concl}.

\section{Continuum solitary waves}
\label{sec:cont}

The double-well potential with the coupling parameter $\lambda$ and
the two minima at $\phi=\pm a$
\be V= \frac{\lambda}{4} \ (\phi^2-a^2)^2
\ee
leads to the following relativistic field equation
\be
   - \frac{\partial^2\phi}{\partial t^2}
   + \frac{\partial^2\phi}{\partial x^2}
   - \lambda \phi(\phi^2-a^2) = 0 \>.
\ee
For $\lambda>0$, if one assumes a moving periodic solution with velocity
$v$ in terms of the Jacobi elliptic function ${\rm sn(x,m)}$ with
modulus $m$
\be
\phi(x,t) = Aa ~{\rm sn}\left[\beta(x+x_0-vt),m\right]
\ee
one finds that
\begin{align}
   - \frac{\partial^2\phi(x,t)}{\partial t^2}  &
   + \frac{\partial^2\phi(x,t)}{\partial x^2}
   \\ \notag &
   = - \beta^2 a~ (1-v^2)~[(1+m)~\sn -2m~\sn^3]
   \>,
\end{align}
where we have suppressed the argument of ${\rm sn}$. Matching terms
proportional to ${\rm sn}$ and ${\rm sn}^3$ we find we have a moving
periodic solution with
\begin{align}
   \beta & = \left[ \frac{\lambda a^2}{(1-v^2) (1+m)} \right]^{1/2} ,
   \nonumber \\
   A & = \sqrt \frac{2m}{1+m} .
\end{align}
The usual kink solitary wave is obtained in the limit
$m~\rightarrow~1$. Since \be
   {\rm sn}(x,1) = \tanh(x) \>,
\ee
we obtain
\be
    \phi(x,t) = a \tanh \Biggl \{  \left[ \frac{\lambda
    a^2}{2(1-v^2)} \right]^{1/2}(x+x_0-vt) \Biggr \} \>.
\ee
If instead $\lambda
<0$, then there are $\mathrm{dn}$ Jacobi elliptic function solutions to
the field equations. Assuming \be
   \phi(x,t) = Aa ~{\dn}\left[\beta(x+x_0-vt),m\right] \>,
\ee
and matching terms proportional to ${\rm dn}$ and ${\rm dn}^3$,
we find a moving periodic pulse solution with
\begin{align}
   \beta & = \left[ \frac{-\lambda a^2}{(1-v^2) (2-m)} \right]^{1/2} ,
   \nonumber \\
   A & = \sqrt \frac{2}{2-m}
   \>.
\end{align}

In fact, there is another pulse solution in terms of the Jacobi elliptic
function of ${\rm cn}$ type.
Assuming \be \phi(x,t) = Aa ~{\cn}\left[\beta(x+x_0-vt),m\right]\>. \ee
Matching terms proportional to ${\rm cn}$ and ${\rm cn}^3$ we find
we have a moving periodic solution with
\begin{align}
   \beta & = \left[ \frac{-\lambda a^2}{(1-v^2) (2m-1)} \right]^{1/2} ,
   \nonumber \\
   A & = \sqrt \frac{2m}{2m-1} .
\end{align}
Note that this solution is valid only if $m > 1/2$.

The usual pulse solitary wave is obtained in the limit $m
\rightarrow 1$. Since \be
   {\rm dn}(x,1) = {\rm cn}(x,1) = \sech(x) \>,
\ee we obtain \be
   \phi(x,t) = a \sqrt{2} \sech
   \left[  \left( \frac{-\lambda a^2}{1-v^2} \right)^{1/2}(x+x_0-vt) \right] \>.
\ee
Note that similar solutions and their stability were considered by
Aubry in the context of structural phase transitions~\cite{aubry}.
In the rest of this paper we will refer to the $\mathrm{sn}$ solutions
as kink-like, and to the $\mathrm{dn}$ and $\mathrm{cn}$ solutions as
pulse-like.

\section{Discretization of $\lambda \phi^4$ field theory}
\label{sec:discrete}

A naive discretization of the field equation above is
\be
   - \ddot \phi_n +
   \frac{1}{\epsilon^2} (\phi_{n+1}+\phi_{n-1}-2\phi_n)
   + \lambda \phi_n (a^2 -\phi_n^2)=0 \>,
\ee
where $\epsilon$ is the lattice parameter and the overdots represent
time derivatives.  However, this equation does not admit solutions
of the form
\be
   \phi_n(t)  =Aa ~ {\sn} (\beta [(n+c)\epsilon-vt],m),
\ee
where $c$ is an arbitrary constant.  To find solutions, one has to
modify the naive discretization.  The key for understanding how to
modify the equation comes from the following identities~\cite{khare}
of the Jacobi elliptic functions

\noindent $\sn$:
\begin{align}
   m~\sn(x,m)^2 & \left[ \sn(x+\beta,m) + \sn(x-\beta,m) \right]
   \\ \notag
   = \ & \ns^2(\beta,m) \bigl [ \sn(x+\beta,m) + \sn(x-\beta,m) \bigr]
   \\ \notag
     & - 2 \cs(\beta,m) \ds(\beta,m) \sn(x,m) \>,
\end{align}
$\cn$:
\begin{align}
   m~\cn(x,m)^2 & \left[ \cn(x+\beta,m) + \cn(x-\beta,m) \right]
   \\ \notag
   = \ & -\ds^2(\beta,m) \bigl [ \cn(x+\beta,m) + \cn(x-\beta,m) \bigr]
   \\ \notag
     & + 2 \cs(\beta,m) \ns(\beta,m) \cn(x,m) \>,
\end{align}
and $\dn$:
\begin{align}
   \dn(x,m)^2 & \left[ \dn(x+\beta,m) + \dn(x-\beta,m) \right]
   \\ \notag
   = \ & - \cs^2(\beta,m) \bigl [ \dn(x+\beta,m) + \dn(x-\beta,m) \bigr ]
   \\ \notag
     & + 2 \ds(\beta,m) \ns(\beta,m) \dn(x,m) \>,
\end{align}
where
\begin{align}
   ns=\frac{1}{\sn(x,m)} \>, \quad
   cs=\frac{\cn(x,m)}{\sn(x,m)} \>, \quad
   ds=\frac{\dn(x,m)}{\sn(x,m)} \>.
\end{align}
It will be useful later to rewrite these identities in the
form:
\begin{align}
\label{eq:sn_id}
   \sn(x+\beta,m) & + \sn(x-\beta,m)
   \\ \notag
   = & \
   2 \ \cs(\beta,m) \ \ds(\beta,m) \ \sn(x,m)
   \\ \notag & \times
   [ \ns^2(\beta,m) -  m \ \sn(x,m)^2] ^{-1} ,
\end{align}
\begin{align}
\label{eq:cn_id}
   \cn(x+\beta,m) & + \cn(x-\beta,m)
   \\ \notag
   = & \
   2 \ \cs(\beta,m) \ \ns(\beta,m) \ \cn(x,m)
   \\ \notag & \times
   [ \ds^2(\beta,m) +  m \ \cn(x,m)^2] ^{-1} ,
\end{align}
and
\begin{align}
\label{eq:dn_id}
   \dn(x+\beta,m) & + \dn(x-\beta,m)
   \\ \notag
   = & \
   2 \ \ds(\beta,m) \ \ns(\beta,m) \ \dn(x,m)
   \\ \notag & \times
   [  \dn(x,m)^2  + \cs^2(\beta,m)  ] ^{-1} .
\end{align}
For the sake of brevity, in what follows we will suppress the
modulus $m$ in the argument of the Jacobi elliptic functions, except
when needed for added clarity.

\subsection{Static lattice solutions}

Consider the naive lattice equation:
\be
   \frac{1}{\epsilon^2}
   \bigl ( \phi_{n+1}+\phi_{n-1}-2\phi_n \bigr )
   + \lambda \phi_n (a^2 - \phi_n^2) = 0 \>.
\ee
Here $\phi_n \equiv \phi[ \beta\epsilon (n+c), m]$.
A general ansatz to modify the solution is to multiply the second difference
operator by the factor $(1-\alpha \phi_n^2)$, with $\alpha$ chosen so that
we get a consistent set of equations.  That is, we consider instead the equation:
\be
   \frac{1- \alpha \phi_n^2}{\epsilon^2}
   \bigl ( \phi_{n+1}+\phi_{n-1}-2\phi_n \bigr )
   +\lambda \phi_n (a^2 - \phi_n^2) =0.  \label{eq:static}
\ee
Using $\alpha$ to eliminate the $\phi_n^3$ term leads to the result
\be
\alpha = \lambda \epsilon^2/2.
\ee
This implies, in the static case, that the lattice equation is just a smeared
discretization of the $\phi_n^3$ term.  Namely, one merely needs to study
the lattice equation:
\be
   \frac{1}{\epsilon^2}
   \bigl ( \phi_{n+1}+\phi_{n-1}-2\phi_n \bigr )
   = \frac{\lambda \phi_n^2}{2}  \bigl ( \phi_{n+1}+\phi_{n-1} \bigr ) - \lambda \phi_n  a^2 .
\ee
Assume a solution of the form (for $\lambda>0$)
\be
\phi_n = Aa~ \sn_n\,,
\ee
where $\sn_n$ denotes $\sn[\beta\epsilon (n+c),m]$ with $c$ being an
arbitrary constant. Note that we only need to consider $c$ between 0
and 1/2 (half the lattice spacing).
On matching the coefficients of $\sn_n$ and $\sn_{n+1}+\sn_{n-1}$ we obtain
\be
A^2 a^2 = \frac{(1+m) ~\sn^2(\beta \epsilon)}{\lambda \epsilon^2} ,
   \label{eq:sn1}
\ee
and
\be
\lambda a^2 = \frac{2}{\epsilon^2} \left[ 1- \dn (\beta \epsilon) \cn (\beta \epsilon)\right] \>.
   \label{eq:sn2}
\ee
In the continuum limit ($\epsilon \rightarrow 0$), we get
\be
\beta^2 = \frac{\lambda a^2}{1+m};  ~~~ A^2= \frac{2m}{1+m} \>,
\ee
which agrees with our continuum result (2.5) with $v=0$.
Also as $m \rightarrow 1$ we recover the usual kink solution
\be
\phi_n =  a \ \tanh [\beta\epsilon (n+c)] \,.
\ee

If instead, we have $\lambda < 0$ and we assume
\be
\phi_n = Aa~ \dn_n,
\ee
we obtain:
\be
A^2 a^2  = - \frac{2m}{\lambda \epsilon^2} \frac{\sn^2(\beta \epsilon)}{\cn^2(\beta \epsilon)} \>,
   \label{eq:dn1}
\ee
and
\be
\lambda a^2 \frac{2}{\epsilon^2} \left[ 1-  \frac{\dn(\beta \epsilon)}{ \cn^2(\beta \epsilon)} \right] \>.
   \label{eq:dn2}
\ee
In the limit when the lattice spacing $\epsilon$ goes to zero, we obtain
\be
\beta^2= - \frac{\lambda a^2}{2-m}; ~~~   A^2 = \frac {2}{2-m},
\ee
same as the continuum case (2.9) with $v=0$.

For $\lambda < 0$ we also have a possible solution of the form
\be
\phi_n = Aa~ \cn_n,
\ee
we obtain:
\be
A^2 a^2  = - \frac{2m}{\lambda \epsilon^2} \frac{\sn^2(\beta \epsilon)}{\dn^2(\beta \epsilon)} ,
   \label{eq:cn1}
\ee
and
\be
\lambda a^2 = \frac{2}{\epsilon^2} \left[ 1-  \frac{\cn(\beta \epsilon)}{ \dn^2(\beta \epsilon)} \right].
   \label{eq:cn2}
\ee
Again taking the lattice spacing to zero, we obtain
\be
  \beta^2= - \frac{\lambda a^2}{2m-1}; ~~~   A^2 = \frac {2m}{2m-1},
\ee
same as the continuum case (2.11) with $v=0$.  Note that as before this
solution is only valid if $m > 1/2$.  For both these solutions, as
$m \rightarrow 1 $, we get our previous continuum solution:
\be
\phi_n =  \sqrt{2} a \ \sech[\beta\epsilon(n+c)] \>.
\ee

\subsection{Stability of solutions}

\begin{figure}[t!]
   \includegraphics[width=0.47\textwidth]{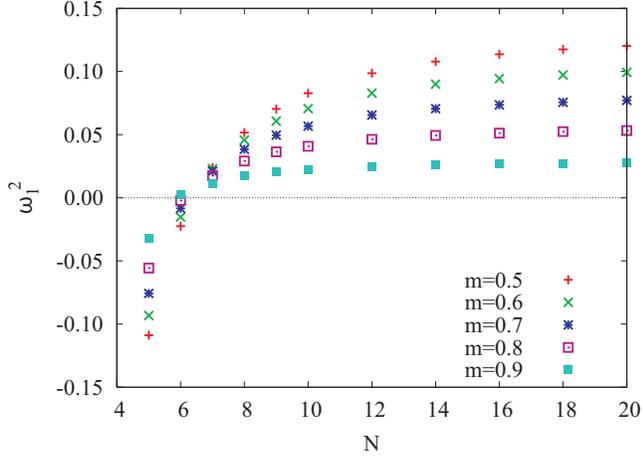}
   \caption{\label{fig:stab_sn}
   (Color online) Kink-like ($\lambda>0$) case: N-dependence of the lowest eigenvalue, $\omega_1^2$, for $\lambda=1,\ a=1$, and various values of the
elliptic modulus $m$.
   }
\end{figure}

\begin{figure}[!]
   \includegraphics[width=0.47\textwidth]{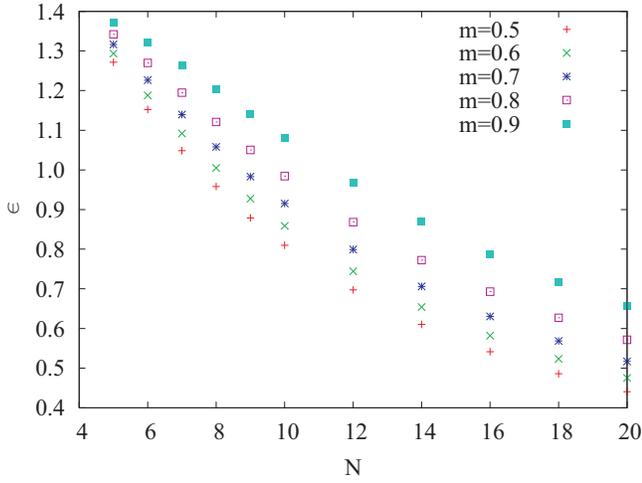}
   \caption{\label{fig:eps_sn}
   (Color online) Kink-like ($\lambda>0$) case: N-dependence of the lattice spacing, $\epsilon$,
   for $\lambda=1,\ a=1$, and various values of $m$.
   }
\end{figure}

\begin{figure}[t!]
   \includegraphics[width=0.47\textwidth]{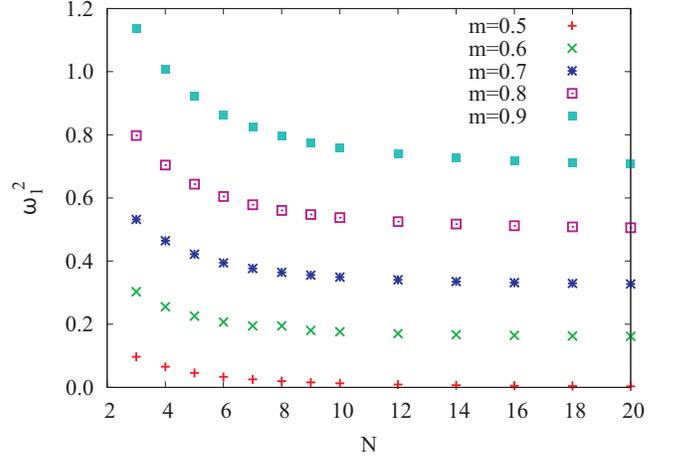}
   \caption{\label{fig:stab_dn}
   (Color online) Pulse-like ($\lambda<0$) $\dn$ case: N-dependence of
the lowest eigenvalue, $\omega_1^2$, for $\lambda=-1,\ a=1.5$, and
various values of $m$.
}
\end{figure}

\begin{figure}[!]
   \includegraphics[width=0.47\textwidth]{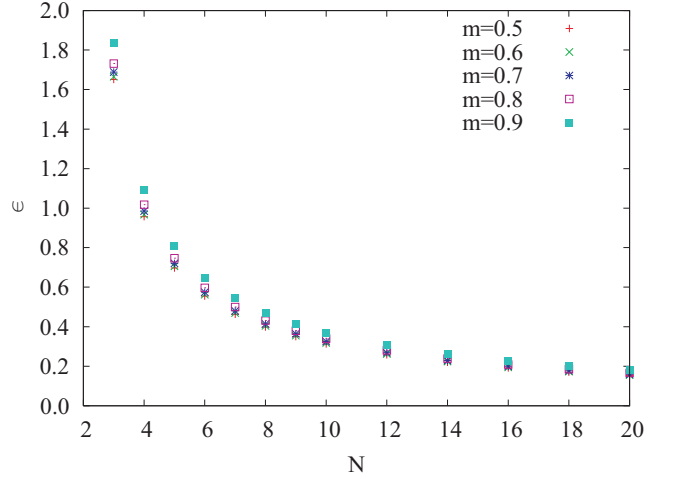}
   \caption{\label{fig:eps_dn}
   (Color online) Pulse-like ($\lambda<0$) $\dn$ case: N-dependence of the lattice spacing, $\epsilon$,
   for $\lambda=-1,\ a=1.5$, and various values of $m$.
   }
\end{figure}

Let us now discuss the stability of these static solutions. To that purpose,
let us expand
$$\phi_n = \phi_n^{(s)} + \psi_n \exp(-i\omega t) , $$
where $\phi_n^{(s)}$ is the known solitary wave exact solution and
$\psi_n$ is a small perturbation.  Then we find that to the lowest order
in $\psi_n$ the stability equation is
\bea  &&\omega^2\psi_n +\frac{1}{\epsilon^2}(\psi_{n+1}
+\psi_{n-1}-2\psi_n)
\nonumber \\ && - \lambda  \ \psi_n \phi_n^{(s)}[\phi^{(s)}_{n+1}+\phi^{(s)}_{n-1}] \nonumber \\
&& +\lambda \psi_n  a^2 - \frac{\lambda \phi_n^2}{2}
(\psi_{n+1}+\psi_{n-1})=0\>. \eea Using the
identities~\eqref{eq:sn_id}, (3.8) and~\eqref{eq:dn_id}, the combination
$[\phi^{(s)}_{n+1}+\phi^{(s)}_{n-1}]$ can be written as a function
of $\phi_n$ alone.  Thus schematically we have
\be \bigl ( \omega^2 - f_n \bigr ) \psi_n
\ + \ g_n \bigl ( \psi_{n+1}+\psi_{n-1} \bigr ) \ = \ 0 \>,
\ee
or
\be
\psi_{n+1}+\psi_{n-1} + h_n \psi_n =0 \>,
\ee
with
\be h_n = \frac{\omega^2 - f_n}{g_n} \>,
\ee
where $f_n$ and $g_n$ are well defined functions.  We also have
that the solutions are periodic on the lattice with $N$ sites ($\psi_{n+N}
= \psi_{n}$). If we write the system of $N$ equations in matrix form, as
\be
\mathbf{A} \ [\psi] \ = \ \omega^2 \ [\psi] ,
\ee
then the condition for nontrivial solutions is that
\be
  \mathrm{Det}\  | \mathbf{A} - \omega^2 \mathbf{1}| = 0.
\ee

In our simulations we require that the solution has exactly one
period in the model space. Therefore, the equations for $\beta$ and
$A$ are supplemented by the equations
   \begin{equation}
      N \beta \epsilon \ = \ 4 \ K(m) \>,
   \label{eq:sn3}
   \end{equation}
for solutions, $\phi_n=\mathrm{sn}(\beta n \epsilon)$ or
$\phi_n=\mathrm{cn} (\beta n \epsilon)$, and
   \begin{equation}
      N \beta \epsilon \ = \ 2 \ K(m) \>,
   \label{eq:dn3}
   \end{equation}
when $\phi_n=\mathrm{dn}(\beta n \epsilon)$.  Here $K(m)$ is the
complete elliptic integral of the first kind \cite{stegun,byrd}.
Hence, once the parameters $m$, $\lambda$ and $a$ are
specified, we need to solve a system of equations for
$\beta$, $\epsilon$, and~$A$, i.e. we have to solve
Eqs.~(\ref{eq:sn1}, \ref{eq:sn2}, \ref{eq:sn3}) or
Eqs.~(\ref{eq:dn1}, \ref{eq:dn2}, \ref{eq:dn3}), or Eqs. (\ref{eq:cn1}, \ref{eq:cn2}, \ref{eq:sn3}),
for the solutions $\mathrm{sn}$, $\mathrm{dn}$, and $\mathrm{cn}$, respectively, for
$\beta$, $\epsilon$, and $A$ as a function of~N.

Typical results are depicted in Figs.~\ref{fig:stab_sn} and
\ref{fig:eps_sn} for kink-like $\sn$ solutions, and in
Figs.~\ref{fig:stab_dn} and \ref{fig:eps_dn}, for pulse-like $\dn$
solutions, respectively. Similar results can also be obtained in the
case of the $\cn$-type pulse solution. For the case of the kink-like
solutions and our choice of parameters, we find from
Fig.~\ref{fig:stab_sn}, that stability requires $N>7$ lattice sites.
In the pulse-like case, we find stability for arbitrary values of
$N$. The magnitude of the lowest eigenvalues increases with $N$ in
the kink-like case, while it decreases with N for pulse-like
solutions. For fixed $m$, $\lambda$, and $A$, the lattice spacing,
$\epsilon$, is not an independent quantity, but it is a well-defined
function of~$N$. As seen from Figs.~\ref{fig:eps_sn} and~\ref{fig:eps_dn},
the lattice spacing, $\epsilon$, is always a decreasing function
of~$N$.

\section{Hamiltonian dynamics}
\label{sec:hdyn}
In this section we demonstrate that our discrete model is a Hamiltonian
system.  The equation which the static solutions obey,
Eq.~(\ref{eq:static}), can be written as:
\be
   \frac{1}{\epsilon^2} \bigl ( \phi_{n+1}+\phi_{n-1}-2\phi_n \bigr )
+\lambda \frac{\phi_n (a^2 - \phi_n^2)}{1- \alpha \phi_n^2} =0.  \label{eq:static2}
\ee
We recognize that this equation can be derived from the potential function
\be
V_0[\phi_n]= \sum_n \frac{(\phi_{n+1}-\phi_n)^2}{2 \epsilon^2}
+ \lambda \sum_n \int d\phi_n \frac{ \phi_n (a^2 - \phi_n^2)}{1- \alpha \phi_n^2}.
\ee
Performing the integral we obtain explicitly:
\begin{align}
   V_0[\phi_n]= & \sum_n \frac{(\phi_{n+1}-\phi_n)^2}{2 \epsilon^2}
   \label{eq:V0} \\ \notag &
   \ - \ \lambda  \frac{ \phi_n^2}{2\alpha}
   \ + \
   \lambda (a^2 \alpha-1) \ \frac{\ln(1- \alpha \phi_n^2)}{2 \alpha^2}.
\end{align}
One easily verifies that
\begin{align}
   - (1- \alpha \phi_n^2) \ \frac{\partial V_0}{\partial \phi_n}
   \ = \ &
   \frac{1- \alpha \phi_n^2}{\epsilon^2}
   \bigl ( \phi_{n+1} +\phi_{n-1}-2\phi_n \bigr )
   \notag \\ &
   \ + \ \lambda \phi_n (a^2 - \phi_n^2) .
\end{align}
The continuum limit $\alpha \rightarrow 0$ (or $\epsilon \rightarrow
0$) of the last two terms in Eq.~\eqref{eq:V0} is given by
\begin{align}
   \frac{\lambda}{4} (\phi_n^2 - a^2)^2 \ - \ \frac{\lambda}{4} a^4
   \>.
\end{align}

We will now show that if we want our class of solutions to be the
static limit of a Hamiltonian dynamical system, that we will be able
to have single solitary waves obey a simple equation, but general
solutions will obey a more complicated dynamics with terms
proportional to ${\dot \phi}^2$. For simplicity we will assume that
the Hamiltonian takes the form: \be H = \sum_n \left( \frac{\pi^2_n}{2}
 \, g[\phi_n] \ + \ V[\phi_n] \right ) \label{eq:H}, \ee
with $V$ given by $V_0$ in Eq.~(\ref{eq:V0}) plus possibly some
additional terms that vanish in the continuum limit.  Here $\pi_n$
is the conjugate momentum and $g[\phi_n]$ a weight function.  For
generality we will assume, as in the case of the discrete nonlinear
Schr\"odinger (DNLS) equation~\cite{dnls,das}, that an extended
Poisson bracket structure exists, namely: \be \left\{ \phi_m, \pi_n
\right\} = \delta_{nm} f[\phi_n] , \ee and
\begin{align}
{\dot \phi}_n = \left\{\phi_n, H \right\} &= \frac {\partial H}{\partial \pi_m} \left\{\phi_n, \pi_m\right\} = f[\phi_n] \frac {\partial H}{\partial \pi_n} ,
\nonumber \\
{\dot \pi}_n = \left\{\pi_n, H\right\} &= \frac {\partial H}{\partial \phi_m}\left\{\pi_n, \phi_m\right\}= -f[\phi_n] \frac {\partial H}{\partial \phi_n} .
\nonumber \\
\end{align}
From our ansatz, Eq.~(\ref{eq:H}), we obtain the first order equations:
\be
{\dot \phi}_n = \pi_n f[\phi_n] g[\phi_n] \equiv \pi_n h[\phi_n] ,
\ee
and
\be
{\dot \pi}_n= - f[\phi_n] \left( \frac{\pi_n^2}{2} \frac{\partial g}{\partial \phi_n} + \frac{\partial V }{\partial \phi_n}   \right) .
\ee
This leads to the following second order differential equation for $\phi_n$
\be
{\ddot \phi}_n = - f[\phi_n] h[\phi_n] \frac{\partial V}{\partial \phi_n} +{\dot \phi}^2 \left(\frac{1}{h} \frac{\partial h}{\partial \phi_n} - \frac{f}{2h} \frac{\partial g}{\partial \phi_n} \right).
\ee
For this equation to have the previously found static solitary waves, as well as the correct
continuum limit, we need only that
\be
f[\phi_n] h[\phi_n] = f^2[\phi_n] g[\phi_n] = 1- \alpha \phi_n^2.
\ee
The two simplest cases are: (a) $f=1$  (ordinary Poisson brackets) and
(b) $h=1$ (extended Poisson brackets).  First consider the case $ f=1$.
This requires
\be
h[\phi_n] = g[\phi_n] = 1- \alpha \phi_n^2 .
\ee
From this we get, if we choose $V=V_0$ with $V_0$ given by
Eq.~(\ref{eq:V0}), the equation:
\begin{align}
   \ddot \phi_n \ = \ &
   \frac{1- \alpha \phi_n^2}{\epsilon^2} \bigl ( \phi_{n+1}+\phi_{n-1}-2\phi_n \bigr )
   \\ \notag &
   \ + \ \lambda \phi_n (a^2 - \phi_n^2)
   \ - \ \frac{\alpha{\dot \phi_n}^2 \phi_n}{1-\alpha\phi_n^2} .
\end{align}
For $\alpha = \lambda {\epsilon^2}/{2}$ this equation has the previously
found static lattice solitary wave solutions as well as the correct
continuum limit.
For the case $h=1$, this leads to
\be
f=  1- \alpha \phi_n^2\>, ~~~  g= (1- \alpha \phi_n^2)^{-1}\>.
\ee
We find that for $ V = V_0$  the equation of motion again becomes
the {\it same} as Eq.~(4.14).


Unfortunately the presence of the ${\dot \phi}^2$ term does not allow
for single elliptic time dependent solitary wave solutions that are of
the form $\sn[\beta(x-ct)]$, $\cn[\beta(x-ct)]$
or $\dn[\beta(x-ct)]$ where $x = n \epsilon$.
This is because the second derivative of the $\sn$ and $\dn$ (or $\cn$)
functions contains both linear as well as cubic terms, and
the quantity ${\dot \phi}^2$ is a quartic polynomial in $\sn$ or
$\dn$ (or $\cn$).  Thus the last term is equivalent to a non-polynomial
potential when applied to a single elliptic function solution.
Therefore in order to obtain a simple elliptic function solution in the
time dependent case, one must add a non-polynomial potential to $V_0$
which is chosen to exactly cancel the last term when evaluated for a
single elliptic solitary wave solution of the form  $\sn[\beta(x-ct)]$,
$\cn[\beta(x-ct)]$ or
$\dn[\beta(x-ct)]$. This means that to obtain an equation that is
derivable from a Lagrangian or a Hamiltonian we should add to the
static potential $V_0$ an additional contribution $\Delta V$ such
that
\be\label{4.16}
(1-\alpha\phi_n^2)\ \frac{\partial\Delta
V}{\partial\phi_n}=-\frac{\alpha\phi_n {\dot
\phi_n}^2}{1-\alpha\phi_n^2} ,
\ee
where $\phi_n$ is a single
solitary wave described by a time translated elliptic function. It
will turn out that $\Delta V$ needed to obtain a simple solution of
the elliptic kind will depend on the velocity squared of the
solitary wave as well as the type of solution (pulse- or kink-like).

In general we have that the Hamiltonian dynamics leads to
\be
{\ddot \phi}_n=-fh\frac{\partial V_0 +\Delta V}{\partial\phi_n}+{\dot \phi_n}^2
\left(\frac{1}{h}\frac{\partial h}{\partial\phi_n}-\frac{f}{2h}\frac{\partial
g}{\partial\phi_n}\right) .
\ee
Since $fh=1-\alpha\phi_n^2$ and $h=fg$, it follows that
\be
\frac{1}{h}\frac{\partial h}{\partial\phi_n}-\frac{f}{2h}\frac{\partial
g}{\partial\phi_n} = -\frac{\alpha\phi_n}{1-\alpha\phi_n^2} ,
\ee
from where the above condition for $\Delta V$ follows readily.  For the single solitary
wave solutions of the $\sn$, $\cn$ and $\dn$ type one  wants to choose
\be\label{4.19}
\frac{\partial\Delta V}{\partial\phi_n}=\frac{a_1\phi_n^5+a_2\phi_n^3+a_3\phi_n}
{(1-\alpha\phi_n^2)^2} ,
\ee
with $a_i$ dependent on the choice of elliptic function, in order to
cancel the effect of the ${\dot \phi}^2$ terms in the corresponding
equation of motion.

Integrating, we get for $\Delta V$
\begin{align}
   \Delta V
   \ = \ & \frac{a_1\phi^2}{2\alpha^2}+\frac{1}{2(1-\alpha\phi^2)}\left(\frac{a_1}
{\alpha^3}+\frac{a_2}{\alpha^2}+\frac{a_3}{\alpha}\right)
   \notag \\ &
\label{4.20}
   +\left(\frac{a_1}
   {\alpha^3}+\frac{a_2}{2\alpha^2}\right) {\rm ln}(1-\alpha\phi^2) .
\end{align}
If we assume a soliton solution of the form
\be \phi_n = Aa ~\sn \Bigl \{ \beta [(n+c)\epsilon-vt] ,m \Bigr \} \>,
\ee
where $c$ is an arbitrary constant, then
\be
{\dot \phi_n}^2=\beta^2v^2(Aa)^2[1-(1+m)\ \sn^2+m\ \sn^4] ,
\ee
which leads to
\begin{align}
   a_1 \ = \ & -\ m\alpha\left(\frac{\beta v}{Aa}\right)^2 ,
   \\
   a_2 \ = \ & \alpha\beta^2 v^2(1+m) ,
   \\
   a_3 \ = \ & -\ \alpha(\beta v Aa)^2 .
\end{align}
Explicitly,  for the $\sn$ type of solitary wave with a kink limit
when $m=1$  we need to choose the extra potential term to satisfy
\be \label{4.23}
-\frac{\partial\Delta
V}{\partial\phi_n}=\frac{\alpha(\beta vAa)^2\phi_n}
{(1-\alpha\phi_n^2)^2}\left[1-\frac{(1+m)\phi_n^2}{(Aa)^2}+\frac{m\phi_n^4}
{(Aa)^4}\right] . \ee

If instead, we assume a pulse soliton solution (for $\lambda~<~0$) of
the form
\be
\phi_n=Aa ~\dn \Bigl \{ \beta [(n+c)\epsilon-vt] ,m \Bigr \} \>,
\ee
then we have
\be
{\dot \phi_n}^2=\beta^2v^2(Aa)^2[(m-1)+(2-m)\dn^2-\dn^4] ,
\ee
which leads to the (different set) of coefficients $\{ a_i \}$ for
the extra term needed to be added to the Hamiltonian
\begin{align}
   a_1 \ = \ & \alpha\left(\frac{\beta v}{Aa}\right)^2 \>,
   \\
   a_2 \ = \ & - \ \alpha\beta^2 v^2(2-m) \>,
   \\
   a_3 \ =\ & (1-m)\alpha({\beta v}{Aa})^2 \>.
\end{align}
Explicitly, for the  case of $\dn$ solitary waves we choose our additional potential term to satisfy:
\begin{align}
\label{4.29}
   - \frac{\partial\Delta V}{\partial\phi_n}
   = &
   \frac{\alpha(\beta vAa)^2\phi_n}{(1-\alpha\phi_n^2)^2}
   \\ \notag & \quad \times
   \left[(m-1)+\frac{(2-m)\phi_n^2}{(Aa)^2}-\frac{\phi_n^4} {(Aa)^4}\right] .
\end{align}

Finally, we assume a pulse solution (for $\lambda < 0$) of the form
\be
\phi_n=Aa ~\cn \Bigl \{ \beta [(n+c) \epsilon-vt +c] ,m \Bigr \} \>,
\ee
then we have
\be
{\dot \phi_n}^2=\beta^2v^2(Aa)^2[(1-m)+(2m-1)\cn^2-m \ \cn^4] ,
\ee
which leads to the (different set) of coefficients $\{ a_i \}$ for
the extra term needed to be added to the Hamiltonian
\begin{align}
   a_1 \ = \ & m\alpha\left(\frac{\beta v}{Aa}\right)^2 \>,
   \\
   a_2 \ = \ & - \ \alpha\beta^2 v^2(2m-1) \>,
   \\
   a_3 \ =\ & -(1-m)\alpha({\beta v}{Aa})^2 \>.
\end{align}
Explicitly, for the  case of $\cn$ solitary waves we choose our
additional potential term to satisfy:
\begin{align}
\label{4.35}
   - \frac{\partial\Delta V}{\partial\phi_n}
   = &
   \frac{\alpha(\beta vAa)^2\phi_n}{(1-\alpha\phi_n^2)^2}
   \\ \notag & \quad \times
   \left[(1-m)+\frac{(2m-1)\phi_n^2}{(Aa)^2}-\frac{m\phi_n^4} {(Aa)^4}\right] .
\end{align}

Because of our ``fine tuning" of $\Delta V$, in all cases, the solitary
waves effectively obey the second order
differential difference equation:
\begin{align}
   - \ddot\phi_n &
   + \frac{1-\alpha \phi_n^2}{\epsilon^2} \bigl ( \phi_{n+1}+\phi_{n-1}-2\phi_n \bigr )
   \notag \\ &
   + \lambda \phi_n (a^2 -\phi_n^2)=0.
\end{align}
As long as $\alpha$  is proportional to $\epsilon^2$, we expect that this
equation has the correct continuum limit.  Next, we demonstrate this for
the three cases explicitly.

\subsection{Positive $\lambda$ and kink solutions}

If we assume a  solution of the form
\be \phi_n = Aa ~\sn \Bigl \{ \beta[ (n+c) \epsilon-vt ] ,m \Bigr \} \>,
\ee
where $c$ is an arbitrary constant we obtain matching terms proportional to
$\bigl [ \sn_{n+1}+\sn_{n-1} \bigr ]$ that
\be
\sn^2(\beta \epsilon) = \alpha \frac{(Aa)^2}{m} \label{eq:plus1} .
\ee
From the term linear in $\sn$ we then obtain
\be
\lambda a^2 = \frac {2}{\epsilon^2} [1- \dn(\beta \epsilon) \cn(\beta \epsilon)] - (1+m) \beta^2 v^2 .
\ee
When $v=0$, we get our previous static solution (3.16).  This equation
becomes in the continuum limit ($\epsilon \rightarrow 0$)
\be
\beta^2 = \frac{\lambda a^2}{(1+m)(1-v^2)} .
\ee
Finally, matching the cubic terms in $\sn$ yields an equation for $A$ namely
\be
   \lambda A^2 a^2 =2 m \frac{ \sn^2(\beta \epsilon)}{\epsilon^2} - 2 m \beta^2 v^2 .
   \label{eq:cubic1}
\ee
From this we deduce that
\be
A^2 =  \frac{ m \left[ \sn^2(\beta \epsilon)-  \beta^2 v^2 \epsilon^2 \right]}
{ [1- \dn(\beta \epsilon) \cn(\beta \epsilon)] - (1+m) \epsilon^2  \beta^2 v^2/2} .
\ee
In the continuum limit we get
\be
A^2 = \frac {2m}{1+m} ,
\ee
which agrees with our continuum result (3.17).
If we divide (\ref{eq:plus1}) by (\ref{eq:cubic1}) we obtain:
\be
\frac{\alpha}{\lambda} = \frac {\sn^2(\beta \epsilon)} {2 [{\sn^2(\beta \epsilon)}/{\epsilon^2} - \beta^2 v^2 ]} .
\ee
For small lattice spacing we get the result
\be
\alpha \rightarrow \frac {\lambda \epsilon^2}{2(1-v^2)} .
\ee
This shows that in order to have a solitary wave of velocity $v$, the
discretization depends on $v^2$.

When $v=0$ this reduces to the static case:
\begin{align*}
   \alpha a^2 A^2 & = m \ \sn^2(\beta \epsilon)\>,
   \\
   \lambda a^2 A^2 & = \frac{2 m \ \sn^2(\beta \epsilon)}{\epsilon^2}\>,
   \\
   \lambda a^2 & = \frac{2}{\epsilon^2} \bigl [ 1-\cn(\epsilon \beta)\dn(\epsilon \beta) \bigr]
   \>,
\end{align*}
which implies that
\begin{align*}
   \alpha a^2 & =1-\cn(\beta \epsilon)\dn(\beta \epsilon)\>,
   \\
   \alpha & = \lambda \epsilon^2 /2 \>,
   \\
   A^2 & = \frac{m \ \sn^2(\beta \epsilon)}{1-\cn(\beta \epsilon)\dn(\beta \epsilon)}
   \>.
\end{align*}

\emph{Localized mode}: Let us now consider the $m \rightarrow 1$ limit
in which case we get a localized kink solution
\be\label{4.71}
\phi_n = Aa \ \tanh[\beta(n+c)\epsilon-vt]\>,
\ee
where
\begin{align}
\label{4.72}
   A=1\>, \quad \alpha a^2 = \tanh^2(\beta \epsilon)\>,
   \\ \notag
   \lambda a^2 =\frac{2}{\epsilon^2} \tanh^2(\beta\epsilon)-2\beta^2 v^2\,.
\end{align}

\subsection{Negative $\lambda$ and pulse solutions}

If we assume a  solution of the form
\be \phi_n = Aa ~\dn \Bigl \{ \beta ([n+c] \epsilon-vt) ,m \Bigr \} \>,
\ee
where $c$ is an arbitrary constant we obtain matching terms proportional to
$\bigl ( \dn_{n+1}+\dn_{n-1}\bigr )$ that
\be
\frac {\sn^2(\beta \epsilon)}{\cn^2(\beta \epsilon)} = - \alpha (Aa)^2.
\label{eq:alpha2}
\ee
From the term linear in $\dn$ we then obtain
\be
-\lambda a^2 = \frac {2}{\epsilon^2}\left[ \frac{\dn(\beta \epsilon)}
{\cn^2(\beta \epsilon)}-1\right] - (2-m) \beta^2 v^2 .
\ee
When $v=0$, we get our previous static solution (3.21). This equation
becomes in
the continuum limit ($\epsilon \rightarrow 0$)
\be
\beta^2 =-  \frac{\lambda a^2}{(2-m)(1-v^2)} .
\ee
Finally, matching the cubic terms in $\dn$ yields an equation for $A$ namely
\be
-\lambda A^2 a^2 =  \frac{2} {\epsilon^2} \frac {\sn^2(\beta \epsilon)}
{\cn^2(\beta \epsilon)} - 2 \beta^2 v^2  \label{eq:lambda2} .
\ee
From this we deduce that
\be
A^2 =  \frac{ \sn^2(\beta \epsilon)-
\beta^2 v^2 \epsilon^2 \cn^2(\beta \epsilon)}
{ \Bigl [\dn(\beta \epsilon)-\cn^2(\beta \epsilon) \Bigr ]
- (2-m) \epsilon^2  \beta^2 v^2 \cn^2(\beta \epsilon) /2} \>.
\ee
In the continuum limit we get
\be
A^2 = \frac {2}{2-m} \>,
\ee
which agrees with our  previous continuum result (3.22).

Dividing (\ref{eq:alpha2}) by (\ref{eq:lambda2}) we obtain
\be
\frac{\alpha}{\lambda} = \frac{\epsilon^2 \sn^2(\beta \epsilon) }
{2  \Bigl[ \sn^2(\beta \epsilon) - \epsilon^2 \beta^2 v^2 \cn^2(\beta \epsilon) \Bigr]} \>.
\ee
For small $\epsilon$ we have
\be
\alpha =\frac{ \lambda \epsilon^2}{ 2 (1-v^2)} .
\ee
When $v \rightarrow 0$, $ \alpha = \lambda \epsilon^2 /2$.  This exactly
cancels the $\lambda \phi_n^3$ term in the equation of motion and we get the
simple discretization for the time independent case.  In the time dependent
case we again  have the unusual result that the discretization needed is a
function of the velocity of the solitary wave.

If instead, we assume a  solution of the form
\be \phi_n = Aa ~\cn \Bigl \{ \beta [(n+c)\epsilon-vt], m \Bigr \} \>,
\ee
where $c$ is an arbitrary constant we obtain matching terms proportional to
$\bigl ( \cn_{n+1}+\cn_{n-1}\bigr )$ that
\be
m \ \frac {\sn^2(\beta \epsilon)}{\dn^2(\beta \epsilon)} = - \alpha (Aa)^2.
\label{eq:alpha21}
\ee
From the term linear in $\cn$ we then obtain
\be
-\lambda a^2 = \frac {2}{\epsilon^2}\left[ \frac{\cn(\beta \epsilon)}
{\dn^2(\beta \epsilon)}-1\right] - (2m-1) \beta^2 v^2 .
\ee
When $v=0$, we get our previous static solution (3.25). This equation
becomes in the continuum limit ($\epsilon \rightarrow 0$)
\be
\beta^2 =-  \frac{\lambda a^2}{(2m-1)(1-v^2)} .
\ee
Finally, matching the cubic terms in $\cn$ yields an equation for $A$ namely
\be
-\lambda A^2 a^2 =  \frac{2m} {\epsilon^2} \frac {\sn^2(\beta \epsilon)}
{\dn^2(\beta \epsilon)} - 2 m \beta^2 v^2  \label{eq:lambda21} .
\ee
From this we deduce that
\be
A^2 =  \frac{ m \Bigl [ \sn^2(\beta \epsilon) - \beta^2 v^2 \epsilon^2 \dn^2(\beta \epsilon) \Bigr ]}
{ \Bigl [\cn(\beta \epsilon)-\dn^2(\beta \epsilon)\Bigr ]  - (2m-1) \epsilon^2  \beta^2 v^2 \dn^2(\beta \epsilon) /2}
\>.
\ee
In the continuum limit we get
\be
A^2 = \frac {2m}{2m-1} \>.
\ee
which agrees with our  previous continuum result (3.26).
Dividing (\ref{eq:alpha21}) by (\ref{eq:lambda21}) we obtain
\be
\frac{\alpha}{\lambda} = \frac{\epsilon^2 \sn^2(\beta \epsilon)}
{2  \Bigl [\sn^2(\beta \epsilon) - \epsilon^2 \beta^2 v^2 \dn^2(\beta \epsilon) \Bigr ]} \>.
\ee
For small $\epsilon$ we have
\be
\alpha =\frac{ \lambda \epsilon^2}{ 2 (1-v^2)} .
\ee
When $v \rightarrow 0$, $ \alpha = \lambda \epsilon^2 /2$.  This exactly
cancels the $\lambda \phi_n^3$ term in the equation of motion and we get the
simple discretization for the time independent case.  In the time dependent
case we again  have the unusual result that the discretization needed is a
function of the velocity of the solitary wave.

\emph{Localized mode}: In the limit of $m \rightarrow 1$, both $\cn$
and $\dn$ solutions reduce to the localized pulse solution
\be\label{4.73}
\phi_n = Aa \ \sech \Bigl \{ \beta[(n+c)\epsilon-vt] \Bigr \} \>,
\ee
where
\begin{align}
\label{4.74}
-\alpha a^2 A^2 & = \sinh^2(\beta\epsilon)\>,
\\ \notag
-\lambda a^2 & =\frac{2}{\epsilon^2} [\cosh(\beta\epsilon)-1] -\beta^2 v^2
\>,
\\ \notag
-\lambda \ a^2 A^2 & \frac{2}{\epsilon^2}\sinh^2(\beta\epsilon)-2\beta^2 v^2\>.
\end{align}

\begin{figure}[t!]
   \includegraphics[height=0.8\textheight]{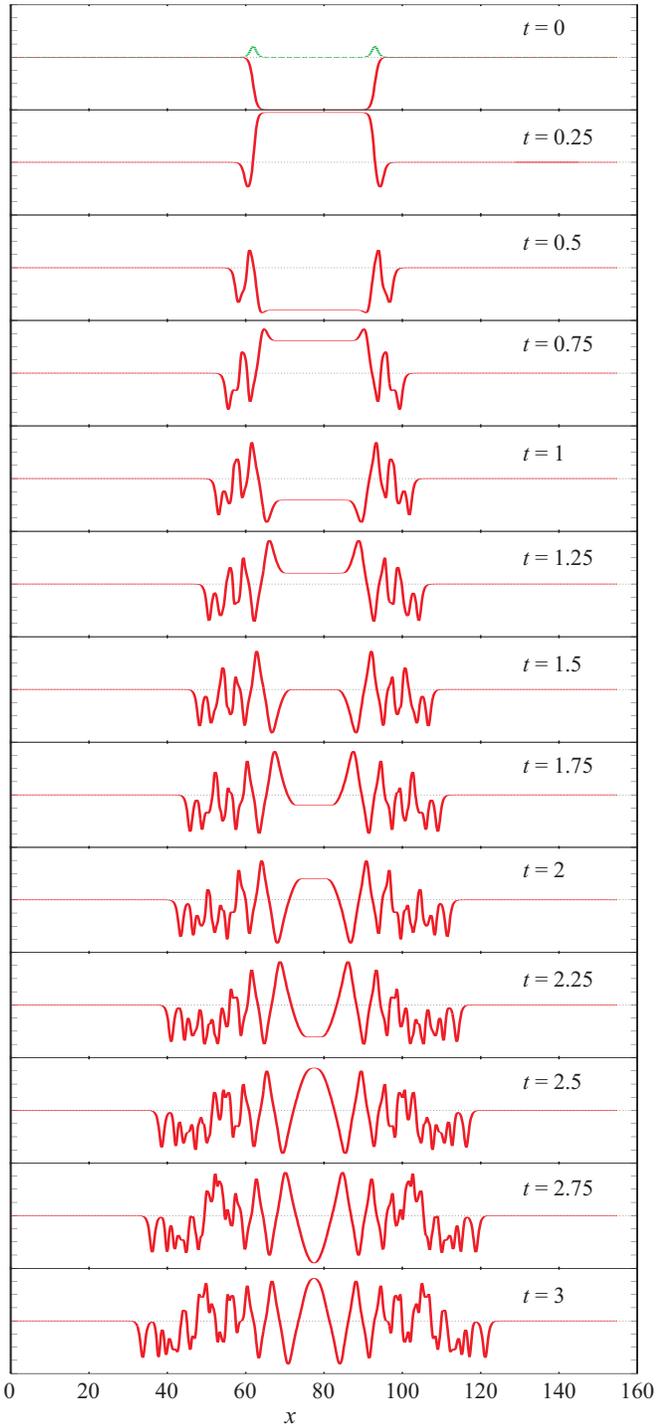}
   \caption{\label{fig:Fig_sn}
   (Color online) Scattering of kink--anti-kink waves. For completeness, at $t=0$, we
   also depict the time-derivative of the initial wave function (in
   green).
   }
\end{figure}

\begin{figure}[t!]
   \includegraphics[height=0.8\textheight]{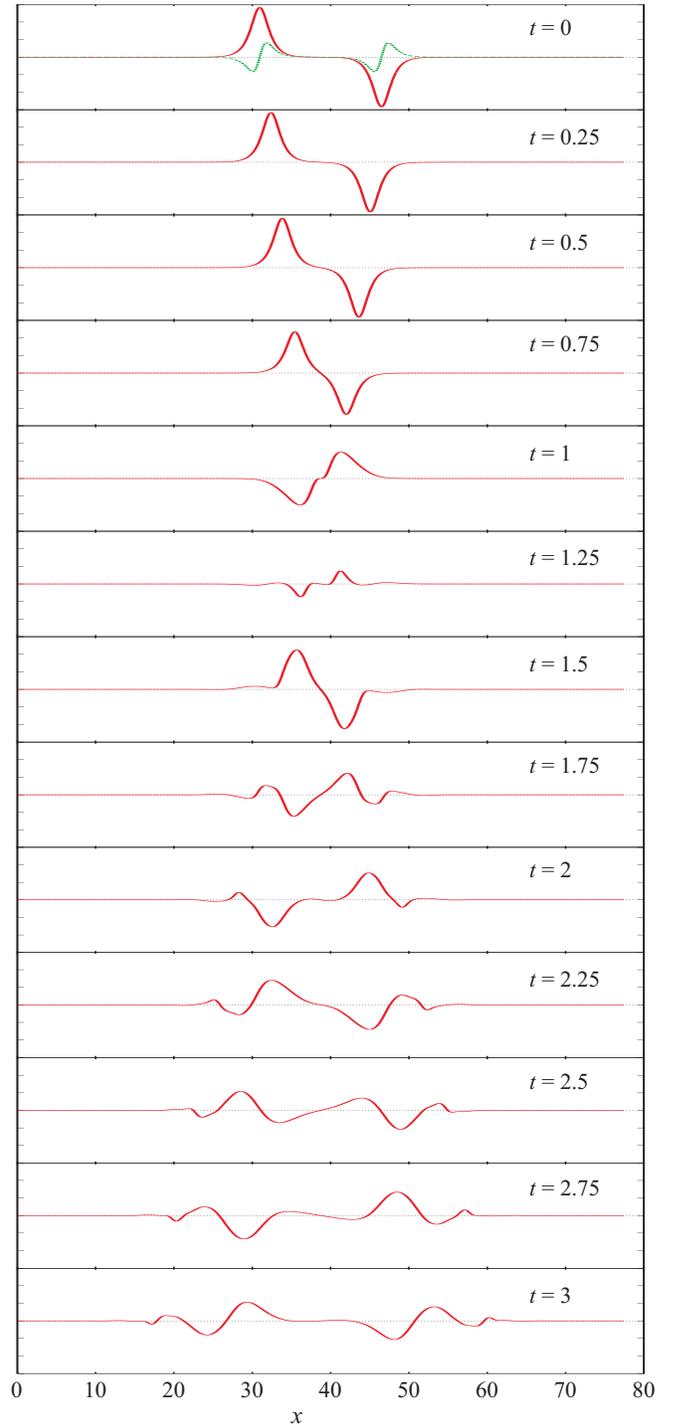}
   \caption{\label{fig:Fig_pm}
   (Color online) Scattering of $\dn$ pulse--anti-pulse waves. For completeness, at $t=0$, we
   also depict the time-derivative of the initial wave function (in
   green).
   }
\end{figure}

\section{Scattering of solitary waves}
\label{sec:scat}

When we have two solitary waves colliding with opposite velocity,
then the equation of motion depends on whether we are considering
$\sn$, $\cn$ or $\dn$ type solitary waves. In general we have
\begin{align}
   {\ddot \phi}_n  = &
   \frac{1- \alpha \phi_n^2}{\epsilon^2} \bigl [ \phi_{n+1}+\phi_{n-1}-2\phi_n \bigr ]
   + \lambda \phi_n (a^2 - \phi_n^2)
   \notag \\ &
   -\frac {\alpha} {1- \alpha \phi_n^2} {\dot \phi_n}^2 \phi_n
   -(1-\alpha \phi_n^2) \frac{\partial \Delta V}{\partial \phi_n}
   \>,
\end{align}
where the partial derivatives of $\Delta V$ are given by
Eqs.~\eqref{4.23}, \eqref{4.29}, and~\eqref{4.35},
for the three types of solutions, respectively. The
$\Delta V$ term given by Eq.~\eqref{4.20}
adds an extra term to the energy conservation
equation.
The conserved energy is given by
\begin{align}\label{5.5}
  E = \sum_n \biggl [ & \frac{\dot \phi_n^2}{2 (1- \alpha \phi_n^2)}
  + \frac{(\phi_{n+1}-\phi_n)^2}{2 \epsilon^2}
  - \frac{\lambda}{2\alpha} \, \phi_n^2
  \notag \\ &
  + \lambda (a^2 \alpha-1) \frac{\ln(1- \alpha \phi_n^2)}{2 \alpha^2} \biggr ]
  \ + \ \Delta V \>.
\end{align}

Typical scenarios are depicted in Figs.~\ref{fig:Fig_sn}
and~\ref{fig:Fig_pm}, for the scattering case of two kink--anti-kink
waves, and two pulse--anti-pulse waves, respectively. The solution
of the kink--anti-kink ``scattering'' waves appears to correspond to
a spatially localized, persistent time-periodic oscillatory
bound-state (or a breather~\cite{breather1,breather2}) with some
radiation (phonons) at late times.  Breathers are intrinsically
dynamic nonlinear excitations and can be viewed as a bound state
of phonons.  The scattering of pulse--anti-pulse is different: there
is a flip after collision and relatively less radiation.

\section{ Energy of solitary waves and the Peierls-Nabarro barrier}
\label{sec:PN}

In a discrete lattice there is an energy cost associated with moving
a localized mode by a half lattice constant, known as
Peierls-Nabarro (PN) barrier~\cite{PN,peyrard}.  In the $
m\rightarrow 1$ limit the elliptic functions become localized and
become either pulses, or kink-like solitary waves. We shall now show
that, rather remarkably, the PN barrier vanishes in this model for
both types of localized modes. In fact we prove an even stronger
result, that the PN barrier is zero for all three periodic
solutions (in terms of $\sn,\cn$ and $\dn$). If the folklore of
nonzero PN barrier being indicative of non-integrability of the
discrete nonlinear system is correct, then \'a l\`a AL lattice, this
could also be an integrable model. It would be worthwhile if one can
prove this explicitly by finding the Lax pair for this model.

We start from the conserved energy expression given by Eq.~(\ref{5.5})
with $\Delta V$ of Eq.~(\ref{4.20}). First of
all, we shall show that in the time-dependent as well as in the static
cases, the conserved energy is given by
\begin{align}
\label{6.1}
  V[\phi_n]    \sum_n \left[
   - \frac{\phi_{n+1}\phi_n}{ \epsilon^2}
 +B\ln (1- \alpha \phi_n^2) +A
   \right]\>,
\end{align}
where the constants A and B vary for each case.
We first note that in view of Eqs. (\ref{4.16}) and (\ref{4.19}), we
have
\begin{align}
\label{6.2}
\frac{\dot \phi_n^2}{2(1-\alpha \phi_n^2)} \ = \ &
-\frac{1}{2\alpha(1-\alpha \phi_n^2)}
\Bigl ( a_1 \phi_n^4 + a_2 \phi_n^2 + a_3 \Bigr )
\>.
\end{align}
Combining this term with $\Delta V$ of Eq.~\eqref{4.20}
in the energy expression~(\ref{5.5}),
we then find that the conserved energy is given by
\begin{align}
\label{6.3}
  E = \sum_{n}
  \biggl [ &
     \left (
     \frac{1}{\epsilon^2}-\frac{\lambda}{2\alpha}
     + \frac{a_1}{\alpha^2}
     \right )\phi_n^2
     - \frac{\phi_{n+1}\phi_n}{\epsilon^2}
  \\ \notag &
  + \frac{1}{2\alpha^2} \Bigl ( a_2+\frac{a_1}{\alpha} \Bigr)
  \\ \notag &
  + \left (\frac{\lambda a^2}{2 \alpha}
           -\frac{\lambda}{2\alpha^2}
           +\frac{a_1}{\alpha^3}
           +\frac{a_2}{2\alpha^2}
  \right )
 \ln(1- \alpha \phi_n^2)
 \biggr ] \>.
\end{align}

Quite remarkably we find that in the case
of all the (i.e. $\sn,\cn$ as well as $\dn$) solutions
\be\label{6.4}
  \frac{1}{\epsilon^2}-\frac{\lambda}{2\alpha}
+\frac{a_1}{\alpha^2} = 0\>,
\ee
where use has been made of relevant equations in Sec.~\ref{sec:hdyn}. As a result
the $\phi_n^2$ term vanishes. Thus the conserved energy takes a
rather simple form as given by Eq.~(\ref{6.1})
in the case of all of our solutions, and the constants $A$ and $B$ are
given by
\begin{align}
\label{6.4a}
  A \ = \ & \frac{1}{2\alpha^2} \ \Bigl ( a_2+\frac{a_1}{\alpha} \Bigr)
  \>,
  \\
\label{6.4b}
  B \ = \ & \frac{\lambda a^2}{2\alpha}
  -\frac{\lambda}{2\alpha^2}
  +\frac{a_1}{\alpha^3}+\frac{a_2}{2\alpha^2}
  \>.
\end{align}
Here $a_1,a_2,a_3$ have different values for the three cases and
are as given in Sec.~\ref{sec:hdyn}. In the special
case of the static solutions, there is a further simplification in
that the constant term also vanishes, since $a_{1,2,3}$ are all
zero in that case.

Let us now calculate the PN barrier in the case of the three
periodic solutions obtained in Sec.~\ref{sec:hdyn} and show that it
vanishes in all the three cases. Before we give the details, let us
explain the key argument. If we look at the conserved energy
expression as given by Eq.~(\ref{6.1}), we find that there are two
$c$-dependent sums involved here. We also observe that in these
expressions, time $t$ and the constant $c$ always come together in
the combination \be\label{6.5} k_1=\beta(c\epsilon-vt)~. \ee
Further, using the recently discovered identities for Jacobi
elliptic functions we explicitly show that for all the solutions the
first sum is $c$-independent. Since the total energy $E$ as given by
Eq.~(\ref{6.1}) is conserved, its value must be independent of time
$t$ and since time $t$ and the constant $c$ always come together in
the combination $k_1$ as given by Eq.~(\ref{6.5}), it then follows
that the second sum must also be $c$-independent and thus there is
no PN barrier for any of our periodic, and hence also the localized
pulse or kink, solutions.

In particular, the following three
cyclic identities~\cite{khare} will allow us to explicitly perform the
first sum in Eq.~(\ref{6.1}):
\begin{align}
   m \ \sn(x) \ \sn(x+a)    - \ & \ns(a)
   \\ \notag & \times \ \left[ Z(x+a) - Z(x) - Z(a) \right]
   \>,
\end{align}
\begin{align}
   m \ \cn(x) \ \cn(x+a) = &
   m \ \cn(a) \\ \notag
   & + \ds(a) \left[
   Z(x+a) - Z(x) - Z(a) \right]
   \>,
\end{align}
and
\begin{align}
   \dn(x) \ \dn(x+a) = & \dn(a)
   \\ \notag &
   + \cs(a) \left[
   Z(x+a) - Z(x) - Z(a) \right]
   \>.
\end{align}
Here $Z(x) \equiv Z(x,m)$ is the Jacobi zeta function \cite{stegun,byrd}.
In addition, we use the fact that
\begin{align}
   \sum_{n=1}^{N} \
   \Bigl \{ Z[\beta\epsilon (n+1)+k_1, m] - Z(n \beta\epsilon +k_1, m) \Bigr \}
   = 0 \>.
\end{align}

Let us now consider the sum as given by first term of Eq.~(\ref{6.1})
for the three cases. We first consider the kink-like solution
which can also be written as
\be\label{6.6}
   \phi_n = Aa \ \sn(n \beta\epsilon+k_1,m) \>.
\ee
Using the identities given above, we obtain
\begin{align}
\label{6.7}
   \sum_{n=1}^{N}  \left[
   -\frac{(\phi_{n+1}\phi_n)}{ \epsilon^2}\right] = - \frac{(A a)^2
   \ns(\beta \epsilon)}{m \epsilon^2} \ N \ Z(\beta \epsilon)
\>.
\end{align}
For the $\dn$ pulse-like case we have instead
\be\label{6.8}
   \phi_n = Aa \ \dn (n \beta \epsilon+k_1,m)
   \>,
\ee
and the first sum in Eq.~(\ref{6.1}) becomes
\begin{align}\label{6.9}
  &
  \sum_{n=1}^{N} \left[ -\frac{\phi_{n+1}\phi_n}{ \epsilon^2}\right]
  \\ \notag &
  = - \frac{N(A a)^2}{\epsilon^2}
  \left[ \dn(\beta \epsilon) - \cs(\beta \epsilon)Z(\beta \epsilon)
  \right] \>.
\end{align}
Finally, for the $\cn$ pulse-like case we have
\be\label{6.10}
   \phi_n = Aa \ \cn(n \beta\epsilon+k_1,m)
   \>.
\ee
In this case, the first sum in Eq.~(\ref{6.1}) becomes
\begin{align}\label{6.10a}
  &- \
  \sum_{n=1}^{N} \frac{\phi_{n+1}\phi_n}{ \epsilon^2}
  = - \frac{N(A a)^2}{\epsilon^2}
  \left[ \frac{1}{m} Z(\beta \epsilon) \ds(\beta \epsilon) - \cn(\beta \epsilon)
  \right] \>.
\end{align}
It is worth noting that all these sums are independent of the constant
$k_1$ and hence $c$. As argued above, since the total energy E as
given by Eq.~(\ref{6.1}) is conserved (and hence time-independent) and
since $t$ and $c$ always appear together in the combination $k_1$ in Eq.~(\ref{6.1})
therefore it follows that the second sum in Eq.~(\ref{6.1})
must also be independent of $k_1$ and hence $c$. We thus have shown
that the PN barrier is zero for the three periodic solutions and hence
also for the localized solutions (which are obtained from them in the
limit $m=1$).

In fact, it is easy to show that for all the three cases, the second
sum (apart from a trivial $k_1$-independent constant) in Eq. (6.1)
is given by
\be\label{6.11}
\sum_{n=1}^{N} \ln
\left [1+\frac{\sn^2(\beta\epsilon)}{\cn^2(\beta\epsilon)}
\dn^2(n\beta\epsilon+k_1) \right ]~.
\ee
In particular, for the solution (\ref{6.6}) the second sum is given by
\be
B\sum_{n=1}^{N} \bigg \{ \ln [\cn^2(\beta\epsilon)] +
\ln \left [ 1+\frac{\sn^2(\beta\epsilon)}{\cn^2(\beta\epsilon)}
\dn^2(n\beta\epsilon+k_1) \right ] \bigg \}~.
\ee
On the other hand, for the solution (\ref{6.8}), the second sum is
given by
\be
B \sum_{n=1}^{N} \ln \Bigl [ 1+\frac{\sn^2(\beta\epsilon)}{\cn^2(\beta\epsilon)}
\dn^2(n\beta\epsilon+k_1) \Bigr ]~.
\ee
Finally, for the solution (\ref{6.10}), the second sum is given by
\be
B\sum_{n=1}^{N} \bigg \{ \ln \frac{\cn^2(\beta\epsilon)}{\dn^2(\beta\epsilon)} +
\ln \left [ 1+\frac{\sn^2(\beta\epsilon)}{\cn^2(\beta\epsilon)}
\dn^2(n\beta\epsilon+k_1)\right ] \bigg \} \>.
\ee

It is worth remarking at this point that by following the above
arguments, it is easily shown that even in the AL model~\cite{AL}
the PN barrier is zero for the periodic $\dn$ and $\cn$ solutions.
In particular, since in  that case the energy is essentially given
by the first sum in Eq.~(\ref{6.1}), hence using Eqs. (\ref{6.9})
and (\ref{6.10a}) it follows that indeed for both $\dn$ and $\cn$
solutions \cite{scott} there is no PN barrier in the AL model.

In general, we do not know how to write the sum in Eq.~(\ref{6.11})
in a closed form. However, for $m=1$ and $N\rightarrow \infty$, the
sum  of
logarithms in Eq.~\eqref{6.3} can be carried out explicitly, and the
energy of the solitary
wave can be given in a closed form.

We proceed as follows: One can show that, for $N \rightarrow \infty$,
the following identity can be derived from the AL equation (see, for instance,
Ref.~\cite{cai}):
\begin{align}
   \sum_{n=-\infty}^{\infty}
   \ln \Bigl \{ 1 + \sinh^2(\rho) \ \sech^2[ \rho(n - x) ] \Bigr \}
   \ = \ \frac{2}{\rho}
   \>.
\end{align}
Then, for kink-like solutions and $m=1$, we have
\begin{align*}
   \ln (1 - & \alpha \phi_n^2)
      \ln [1 - \tanh^2(\beta \epsilon) \tanh^2(n\beta \epsilon + k_1) ]
   \\
   = \ &
   \ln \sech^2(\beta \epsilon) \ + \
   \ln [ 1 + \sinh^2(\beta \epsilon) \ \sech^2n\beta \epsilon + k_1) ]
   \>.
\end{align*}
Thus, in the limit when $N \rightarrow \infty$ and $m=1$, the sum of
logarithms in Eq.~\eqref{6.3} becomes
\begin{align}
   \frac{2}{\beta \epsilon}
   \ + \
   \sum_n \
   \ln \sech^2(\beta \epsilon)
   \>.
\end{align}
Similarly, in the case of the two pulse solutions,
$\dn$ and $\cn$ are identical for $m = 1$, and we can write
\begin{align*}
   \ln (1 - & \alpha \phi_n^2)
      \ln [1 + \sinh^2(\beta \epsilon) \ \sech^2(n\beta \epsilon + k_1) ]
   \>.
\end{align*}
In the limit when $N \rightarrow \infty$, the sum of
logarithms in Eq.~\eqref{6.3} for the two
pulse solutions, $\dn$ and $\cn$, for $m~=~1$, is simply equal to
$2/(\beta \epsilon)$.

For the case of the kink solution, for $m=1$, $B$ as given by
Eq.~(\ref{6.4b}) simplifies to
\begin{equation}
B=-\frac{\sech^2 (\beta\epsilon)}{\alpha\epsilon^2}\>,
\end{equation}
whereas for the case of the pulse solutions, at $m=1$, $B$ simplifies
to
\begin{equation}
B=-\frac{\cosh (\beta\epsilon)}{\alpha\epsilon^2}\>.
\end{equation}

\section{Conclusions}
\label{sec:concl}

In this paper we have shown how to modify the naive discretization
of $\lambda \phi^4 $ field theory so that the discrete theory is a
Hamiltonian dynamical system containing both static and moving
solitary waves.  We have found three different periodic elliptic
solutions.  For time-dependent solutions we found the unusual result
that the discretization is dependent on the velocity.

In the static case, we have studied the stability of both kink-like
and pulse-like solutions, and have found different qualitative
behavior of the lowest eigenvalue of the stability matrix in the two
cases. For typical values of the model parameters, in the case of
kink-like solutions, we found that stability requires the number of
sites, $N$, to be larger than a minimum value, while for pulse-like
solutions stability is achieved for arbitrary values of $N$. The
magnitude of the lowest eigenvalues increases with $N$ in the
kink-like case, and decreases with $N$ for pulse-like solutions.
The lattice spacing, $\epsilon$, is not an independent parameter
and always decreases with $N$.

We also determined the energy of the solitary wave in the three
cases. Using the Hamiltonian structure, we were able to argue that
the PN barrier~\cite{PN,peyrard} for all solitary waves is zero.
This leads to the intriguing question whether this Hamiltonian
system is integrable.  As an additional result we explicitly showed
that for the two elliptic solutions ($\dn$ and $\cn$) \cite{scott}
of the integrable Ablowitz-Ladik model \cite{AL} the PN barrier is
zero--as one would expect.

The single solitary wave solutions were found to be stable and when
we scattered two such single-kink solitary waves there were two
different behaviors. For pulses, the pulse--anti-pulse solution
leads to scattering with a flip and a little radiation (phonons). The
existence of phonons brings into question the integrability of this
model, in spite of the zero PN barrier. For kink-like solutions we
found a breather-like behavior~\cite{breather2} during the collision.
However, we have \emph{not} found \emph{exact} two-solitary wave or
breather solutions, which would help clarify the integrable nature
of this system.

The results presented here are useful for structural phase
transitions~\cite{dphi4,aubry} and possibly for certain field theoretic
contexts~\cite{luis}.  Our results also hold promise for appropriate
discretizations of other discrete nonlinear soliton-bearing
equations~\cite{clark,scott}.  Possible extension to discrete
integrable models in $2+1$ dimensions, e.g. Kadomtsev-Petviashvili
(KP) hierarchies \cite{KP}, would be especially desirable. Extension
to time-discrete integrable models \cite{capel} is another interesting
possibility.

\section{Acknowledgment} This work was supported in part by the U.S.
National Science Foundation and in part by the Department of Energy.
FC and BM would like to thank the Santa Fe Institute for hospitality.
AK acknowledges the hospitality of the Theoretical Division at LANL
where this work was initiated.


\end{document}